%% file: main.tex
\documentclass[letterpaper]{article} 
\usepackage{aaai25}  
\usepackage{times}  
\usepackage{helvet}  
\usepackage{courier}  
\usepackage[hyphens]{url}  
\usepackage{graphicx} 
\usepackage{natbib}  
\usepackage{caption} 
\usepackage{amsmath}
\usepackage{amssymb}
\usepackage{booktabs}
\usepackage{tabularx}
\usepackage{subfig} 
\usepackage{xcolor}

\title{How Diplomacy Reshapes Online Discourse: \\
Asymmetric Persistence in Online Framing of North Korea}

\newcommand{\equalcontrib}{\thanks{Equal Contribution.}}

\author{
  Hunjun Shin,$^{\dagger}$\equalcontrib
  \quad Hoonbae Moon,$^{\ddagger}$\footnotemark[\value{footnote}]
  \quad Mohit Singhal$^{\dagger}$
}

\affiliations{
    $^{\dagger}$Northeastern University, Boston, MA, USA \\
    $^{\ddagger}$Korea National Defense University, Republic of Korea \\
    \texttt{\{shin.hu, m.singhal\}@northeastern.edu, ansgnsqo1@gmail.com} \\
}

\begin{document}

\maketitle

\begin{abstract}
\input{sections/abstract}
\end{abstract}

\begin{links}
    \link{Code \& Data}{https://github.com/Shinhunjun/Asymmetric-Persistence-in-Online-Framing-of-North-Korea}
\end{links}

\input{sections/introduction}

\input{sections/related_work}
\input{sections/method}
\input{sections/results}

\input{sections/discussion}
\input{sections/conclusion}

\bibliography{references}

\section{Ethics Checklist}
\begin{enumerate}
\item For most authors...
\begin{enumerate}
    \item  Would answering this research question advance science without violating social contracts, such as violating privacy norms, perpetuating unfair profiling, exacerbating the socio-economic divide, or implying disrespect to societies or cultures?
    \textcolor{blue} {Yes}
  \item Do your main claims in the abstract and introduction accurately reflect the paper's contributions and scope?
    \textcolor{blue} {Yes}
   \item Do you clarify how the proposed methodological approach is appropriate for the claims made? 
    \textcolor{blue} {Yes}
   \item Do you clarify what are possible artifacts in the data used, given population-specific distributions?
    \textcolor{green} {NA}
  \item Did you describe the limitations of your work?
    \textcolor{blue} {Yes}
  \item Did you discuss any potential negative societal impacts of your work?
    \textcolor{blue} {No, because we ideate on positive inference from our findings. }
      \item Did you discuss any potential misuse of your work?
    \textcolor{green} {NA}
    \item Did you describe steps taken to prevent or mitigate potential negative outcomes of the research, such as data and model documentation, data anonymization, responsible release, access control, and the reproducibility of findings?
    \textcolor{blue} {Yes}
  \item Have you read the ethics review guidelines and ensured that your paper conforms to them?
   \textcolor{blue} {Yes}
\end{enumerate}

\item Additionally, if your study involves hypotheses testing... \textcolor{green} {NA. Our study does not involve hypotheses testing.}

\begin{enumerate}
  \item Did you clearly state the assumptions underlying all theoretical results?
    \textcolor{green} {NA}
  \item Have you provided justifications for all theoretical results?
    \textcolor{green} {NA}
  \item Did you discuss competing hypotheses or theories that might challenge or complement your theoretical results?
    \textcolor{green} {NA}
  \item Have you considered alternative mechanisms or explanations that might account for the same outcomes observed in your study?
    \textcolor{green} {NA}
  \item Did you address potential biases or limitations in your theoretical framework?
   \textcolor{green} {NA}
  \item Have you related your theoretical results to the existing literature in social science?
    \textcolor{green} {NA}
  \item Did you discuss the implications of your theoretical results for policy, practice, or further research in the social science domain?
    \textcolor{green} {NA}
\end{enumerate}

\item Additionally, if you are including theoretical proofs... \textcolor{green} {NA. Our study does not include theoretical proofs.}

\begin{enumerate}
  \item Did you state the full set of assumptions of all theoretical results?
    \textcolor{green} {NA}
	\item Did you include complete proofs of all theoretical results?
    \textcolor{green} {NA}
\end{enumerate}

\item Additionally, if you ran machine learning experiments...
\begin{enumerate}
  \item Did you include the code, data, and instructions needed to reproduce the main experimental results (either in the supplemental material or as a URL)?
    \textcolor{blue} {Yes. We include code and anonymized data via an anonymous repository link. \url{https://anonymous.4open.science/r/Asymmetric-Persistence-in-Online-Framing-of-North-Korea-F0F7}.}
  \item Did you specify all the training details (e.g., data splits, hyperparameters, how they were chosen)?
    \textcolor{green} {NA. We use a pre-trained LLM (GPT-4o-mini) in a zero-shot setting.}
     \item Did you report error bars (e.g., with respect to the random seed after running experiments multiple times)?
    \textcolor{blue} {Yes. We report 95\% confidence intervals for all DiD estimates.}
	\item Did you include the total amount of compute and the type of resources used (e.g., type of GPUs, internal cluster, or cloud provider)?
    \textcolor{blue} {Yes. We specify the use of OpenAI API (GPT-4o-mini) services.}
     \item Do you justify how the proposed evaluation is sufficient and appropriate to the claims made? 
    \textcolor{blue} {Yes. We validate our LLM classification against human expert annotations.}
     \item Do you discuss what is ``the cost`` of misclassification and fault (in)tolerance?
    \textcolor{green} {NA}
  
\end{enumerate}

\item Additionally, if you are using existing assets (e.g., code, data, models) or curating/releasing new assets, \textbf{without compromising anonymity}...
\begin{enumerate}
  \item If your work uses existing assets, did you cite the creators?
    \textcolor{blue} {Yes}
  \item Did you mention the license of the assets?
    \textcolor{green} {NA}
  \item Did you include any new assets in the supplemental material or as a URL?
    \textcolor{blue} {Yes. We include an anonymous GitHub link to the dataset in this submission. \url{https://anonymous.4open.science/r/Asymmetric-Persistence-in-Online-Framing-of-North-Korea-F0F7}.}
  \item Did you discuss whether and how consent was obtained from people whose data you're using/curating?
    \textcolor{green} {NA}
  \item Did you discuss whether the data you are using/curating contains personally identifiable information or offensive content?
    \textcolor{blue} {Yes}
\item If you are curating or releasing new datasets, did you discuss how you intend to make your datasets FAIR?
\textcolor{blue} {Yes, we are committed to abide by FAIR principles when sharing our dataset upon paper acceptance.}
\item If you are curating or releasing new datasets, did you create a Datasheet for the Dataset? 
\textcolor{blue} {Yes, we commit to creating a Datasheet when sharing our dataset upon paper acceptance.}
\end{enumerate}

\item Additionally, if you used crowdsourcing or conducted research with human subjects, \textbf{without compromising anonymity}... \textcolor{green} {NA. Our study does not involve research with human subjects/participants.}

\begin{enumerate}
  \item Did you include the full text of instructions given to participants and screenshots?
    \textcolor{green} {NA}
  \item Did you describe any potential participant risks, with mentions of Institutional Review Board (IRB) approvals?
    \textcolor{green} {NA}
  \item Did you include the estimated hourly wage paid to participants and the total amount spent on participant compensation?
    \textcolor{green} {NA}
   \item Did you discuss how data is stored, shared, and deidentified?
   \textcolor{green} {NA}
\end{enumerate}

\end{enumerate}

\input{sections/appendix}

\end{document}

%% file: sections/abstract.tex
Public opinion toward foreign adversaries shapes and constrains diplomatic options. Prior research has largely relied on sentiment analysis and survey-based measures, providing limited insight into how sustained narrative changes (beyond transient emotional reactions) might follow diplomatic engagement. This study examines the extent to which high-stakes diplomatic summits shape how adversaries are framed in online discourse.
We analyze U.S.--North Korea summit diplomacy (2018--2019) using a Difference-in-Difference (DiD) design on Reddit discussions. Using multiple control groups (China, Iran, Russia) to adjust for concurrent geopolitical shocks, we integrate a validated \textit{Codebook LLM} framework for framing classification with graph-based discourse network analysis that examines both edge-level relationships and community-level narrative structures. Our results reveal \textit{short-term asymmetric persistence} in framing responses to diplomacy. While both post-level and comment-level sentiment proved transient (improving during the Singapore Summit but fully reverting after the Hanoi failure), framing exhibited significant stability: the shift from threat-oriented to diplomacy-oriented framing was only partially reversed. Structurally, the proportion of threat-oriented edges decreased substantially (48\% $\rightarrow$ 28\%) while diplomacy-oriented structures expanded, and these shifts resisted complete reversion after diplomatic failure. These findings suggest that diplomatic success can leave a short-term but lasting imprint on how adversaries are framed in online discourse, even when subsequent negotiations fail.

%% file: sections/introduction.tex
\section{Introduction}
Public opinion toward foreign adversaries is shaped not only by objective events but also by how those events are discussed and interpreted in public discourse~\cite{baum2015war,entman2004}.
In particular, high-stakes diplomatic summits function as salient narrative moments, drawing widespread attention and prompting shifts in how adversary states are evaluated, debated, and framed online~\cite{wapo2018historic}.
Understanding whether such events produce lasting changes in public discourse or are merely short-lived emotional reactions remains an open empirical question.
Prior research on public opinion and foreign policy has largely focused on survey-based measures of approval or sentiment, often emphasizing immediate affective responses to international crises~\cite{kertzer2017bottom, tomz2020public}.
While prior work captures important emotional dynamics, it provides limited insight into how \textit{narrative structures} evolve over time in participatory online environments, where discourse is not only emotional but also relational and structural. Focusing exclusively on sentiment, consequently, risks overlooking broader discourse dynamics.

Recent work in computational social science has begun addressing this gap by applying quasi-experimental designs to social media data, enabling stronger causal inference about the effects of political events~\cite{kumarswamy2025,horta2023deplatforming,trujillo2022make,chandrasekharan2017you}. We extend this methodological advance to the domain of international relations. 
In this study, we argue that evaluating the public impact of diplomacy requires examining both \textit{content} and \textit{structure} in online discourse. We distinguish between two complementary dimensions: (1) content-level change, reflected in sentiment and framing, and (2) structural change, reflected in how narratives are interconnected within discourse networks. Sentiment captures affective valence, whereas framing captures how an adversary is positioned (such as a military threat or a negotiation partner) within public narratives. These dimensions need not evolve in tandem, and disentangling them is crucial for understanding the broader consequences of diplomatic engagement.

We examine these dynamics through the case of U.S.-North Korea summit diplomacy between 2017 and 2019. The Singapore Summit marked an unprecedented moment of diplomatic engagement between the two countries, while the subsequent Hanoi Summit ended without agreement.
Together, these events provide a natural contrast between diplomatic success and failure, allowing us to examine not only immediate shifts in discourse but also whether earlier narrative changes persist following subsequent setbacks. While our causal identification is defined at the post level, we further examine whether the observed shifts are reflected in highly visible audience responses, which play a key role in participatory discourse on Reddit. To the best of our knowledge, this is the first study to apply causal inference methods to this domain. Specifically, we analyze how diplomatic events reshape both content-level framing and structural organization of online discourse toward an adversary nation.
In particular, we investigate the following research questions:
\begin{itemize}
    \item \textbf{RQ1a (Content Change)}: How do high-stakes diplomatic summits causally affect sentiment and framing toward North Korea in online discourse?
    \item \textbf{RQ1b (Asymmetric Persistence)}: Do the effects of successful diplomacy exhibit asymmetric persistence, such that subsequent diplomatic failure does not fully reverse earlier gains?
    \item \textbf{RQ2 (Structural Reorganization)}: How do these diplomatic events restructure the organization of online discourse, as reflected in changes to discourse networks and narrative connectivity?
    \item \textbf{RQ3 (Audience Propagation)}: Do framing shifts observed in agenda-setting posts propagate to highly visible audience responses (comments)?
    \item \textbf{RQ4 (Methodological Validity)}: To what extent can LLM-based framing classification align with human expert judgments in geopolitical discourse analysis?
\end{itemize}
We analyze Reddit discussions from 2017--2019 using a quasi-experimental analysis, called the Difference-in-Difference (DiD)~\cite{abadie2005semiparametric} design with multiple control countries to isolate the effects of summit diplomacy.
To measure discourse at scale, we use a validated Codebook LLM framework
for framing classification and integrate it with graph-based network analysis~\cite{edge2024graphrag} to capture structural reorganization in narrative connectivity. 

Our key findings indicate that: (1) framing shifted significantly from threat-oriented to diplomacy-oriented following the Singapore Summit, while post-level sentiment effects were transient; and (2) after the Hanoi failure, only 39\% of the framing gains reverted, a pattern consistent with asymmetric persistence. Beyond this case, our results suggest how high-salience diplomatic events can durably reorganize narrative structures in participatory online discourse.
By integrating content-level and structural analyses, this study contributes to computational social science in two primary ways:
\begin{enumerate}
    \item \textbf{Causal Distinction between Sentiment and Framing:} We provide causal evidence that diplomatic summits significantly reshape how a foreign adversary is framed, illustrating dynamics that are distinct from transient sentiment responses.
    \item \textbf{Asymmetric Persistence:} We identify a pattern where framing changes are only partially reversed after diplomatic failure, whereas sentiment fully reverts (a phenomenon we term asymmetric persistence).
\end{enumerate}
Additionally, we introduce a methodological framework that integrates causal inference with LLM-based measurement and graph-based discourse analysis, and validates these findings using audience response data.

%% file: sections/related_work.tex
\section{Related Work} 
\textbf{Framing Dynamics in Diplomatic Discourse.} Framing theory posits that public understanding is shaped by selective presentation~\cite{entman1993, entman2004, iyengar1991}. While foundational work examined static frames in traditional media~\cite{semetko2000}, these categories may not capture the dynamic, participatory nature of online discourse. Recent computational studies address this by examining temporal spillovers~\cite{russo2023spillover}, structural changes~\cite{trujillo2022make}, and network stability~\cite{leifeld2016policy, vandenhole2025discourse}. Previous research further highlights how shocks shift online language~\cite{olteanu2018extremist, rizoiu2018debatenight}, how geopolitical conflicts sustain cross-platform narratives~\cite{kloo2024crossplatform, zhu2024partisan}, and the importance of distinguishing emotional valence from substantive stance~\cite{bestvater2023sentiment}. We extend this literature by measuring how diplomatic events reshape the narrative positioning of an adversary over time, addressing the lack of systematic understanding regarding how ``dialogue'' frames compete with entrenched ``threat'' narratives in participatory environments.

\textbf{Causal Analysis of Social Media Discourse.} Social media platforms are central venues for international political deliberation~\cite{baumgartner2020}. While existing work often relies on correlational analyses of sentiment~\cite{kim2021nktwitter}, computational social science increasingly adopts quasi-experimental designs to allow causal interpretation. Researchers have utilized Difference-in-Difference (DiD)~\cite{russo2023spillover, trujillo2022make} to isolate the effects of platform interventions on user behavior and toxicity~\cite{kumarswamy2025, horta2023deplatforming}. We adapt these causal inference approaches to international relations. By employing a DiD design with multiple control countries (China, Iran, Russia), we isolate the specific impact of the Singapore and Hanoi summits on public discourse, distinguishing event-driven effects from general temporal trends.

\textbf{Large Language Models in Political Science.} LLMs enable scalable political text analysis~\cite{ziems2024}, with recent work demonstrating that models like ChatGPT can classify discourse with zero-shot performance comparable to crowd-workers~\cite{gilardi2023, tornberg2023}. However, applying LLMs to specialized geopolitical domains requires scrutiny, as generic benchmarks may overlook complex diplomatic nuances or reflect underlying biases~\cite{sap2022annotators}. We address this challenge by embedding rigorous expert-defined codebook criteria directly into the LLM prompt and validating predictions against a gold-standard benchmark of domain-expert consensus, thereby aligning automated classification with domain-specific standards.

%% file: sections/method.tex
\section{Method}
\subsection{Data Collection}
\textbf{Agenda-Setting Posts (Primary Data):} We collected Reddit posts from January 2017 to June 2019 using the Arctic Shift API~\cite{arcticshift2024}, targeting eight major subreddits: r/worldnews, r/politics, r/news, r/geopolitics, r/korea, r/northkorea, r/AskAnAmerican, and r/NeutralPolitics. To ensure comprehensive coverage, we employed a keyword-based retrieval strategy, querying for posts containing country-specific keywords (e.g., ``North Korea'', ``China'', ``Iran'', ``Russia'') to construct our treatment and control groups. All posts were filtered to English. As detailed in Table~\ref{tab:dataset}, our final primary dataset consists of 29,688 posts, comprising 10,448 North Korea-related posts (treatment) and 19,240 posts across three control groups.

\textbf{Audience Response Data:} While our primary causal analysis focuses on posts ($N=29,688$), we also collected the full comment trees (root comments and all nested replies) for the top-5 most upvoted root comments from each post. This approach captures both highly visible audience responses ($N=255,391$ total comments across $\sim$52,000 root comments) and the cascading discussions they trigger, representing the most influential layer of participatory discourse.

\textbf{Analytical Role of Comments:} We treat audience comments differently across analytical stages. In content-level analyses, comments are used to assess whether post-level effects are reflected in highly visible, socially reinforced audience interpretations. Accordingly, comment-level analyses do not constitute independent causal estimates. In contrast, in network analyses, comments are examined separately to capture interactional and participatory discourse structures that are distinct from agenda-setting posts.

\textbf{Treatment and Control Groups:} To isolate the causal impact of summit diplomacy, it is essential to control for concurrent geopolitical developments that could confound inference. Relying solely on longitudinal observations of the treatment group (North Korea) risks misinterpreting broader fluctuations in Reddit users' foreign policy sentiment as specific effects of the summits.

\begin{table}[t]
\centering
\setlength{\tabcolsep}{3pt}
\caption{Dataset Overview}
\resizebox{0.75\linewidth}{!}{
\begin{tabular}{llrr}
\hline
\textbf{Group} & \textbf{Topic} & \textbf{Posts} & \textbf{Comments} \\
\hline
Treatment & N. Korea & 10,448 & 70,879 \\
Control 1 & China & 5,921 & 62,057 \\
Control 2 & Iran & 4,749 & 40,572 \\
Control 3 & Russia & 8,570 & 81,883 \\
\hline
\textbf{Total} & & \textbf{29,688} & \textbf{255,391} \\
\hline
\end{tabular}
\label{tab:dataset}
}
\end{table}
We selected control countries exposed to contemporaneous high-salience geopolitical events but not subject to U.S.--North Korea summit diplomacy: China (trade conflict), Iran (post-JCPOA tensions), and Russia (U.S. election interference).
Using multiple controls allows us to assess robustness across different confounding structures. However, the validity of a DiD design rests on the parallel trends assumption: treated and control groups must exhibit similar trends in the outcome variable prior to the intervention. We verified this assumption for all groups (see Appendix Tables~\ref{tab:app_pt_sentiment} and~\ref{tab:app_pt_framing}). For sentiment, all three control groups satisfied parallel trends ($p > 0.05$ for all comparisons). For framing, China and Iran satisfied parallel trends, while Russia exhibited a significant violation for P1$\rightarrow$P2 ($p = 0.01$), likely reflecting differential political attention to Russia during the pre-treatment period, hence, we exclude Russia from framing analyses for the Singapore Summit effect and retain China and Iran as primary counterfactuals.

\begin{figure*}[t]
\centering
\includegraphics[width=0.9\textwidth]{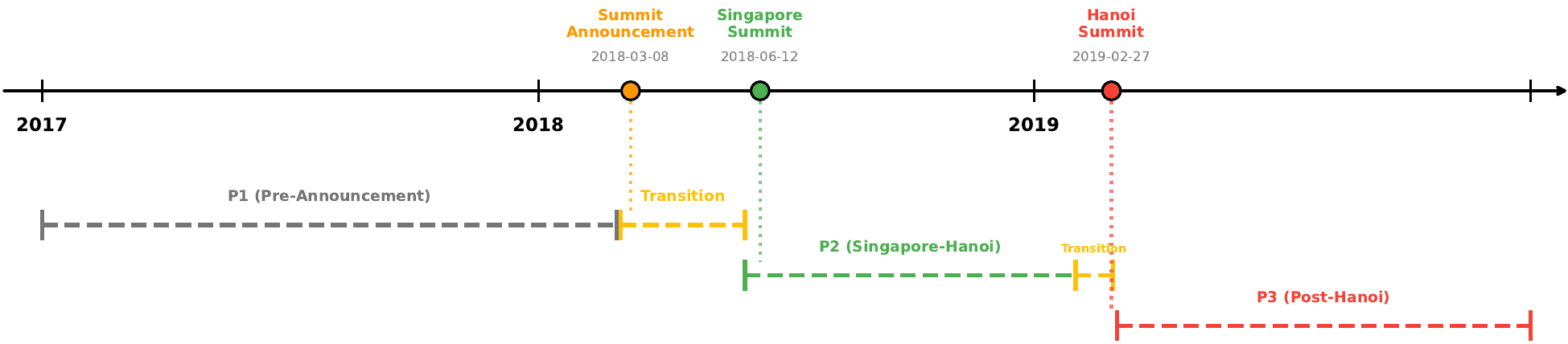}
\caption{Research Timeline and Key Events. Dashed brackets indicate analysis periods: P1 (Pre-Announcement), P2 (Singapore-Hanoi), P3 (Post-Hanoi). Transition periods are excluded from analysis to avoid anticipation effects.}
\label{fig:timeline}
\end{figure*}

\textbf{Key Events and Analysis Periods:} To avoid anticipation effects, we define three analysis periods excluding transition months, and Figure~\ref{fig:timeline} visually summarizes these periods and events:
\begin{itemize}
    \item \textbf{P1} (2017.01--2018.02): Pre-Announcement
    \item \textbf{P2} (2018.06--2019.01): Singapore-Hanoi
    \item \textbf{P3} (2019.03--2019.12): Post-Hanoi
    \item \textit{Excluded}: 2018.03--05, 2019.02 (Transition periods)
\end{itemize}

\subsection{Outcome Measurement}
To answer our research questions, we measure two complementary dimensions of discourse: \textit{sentiment} (emotional valence) and \textit{framing} (how North Korea is narratively positioned). 

\textbf{Sentiment Analysis: } We use the TweetEval RoBERTa model~\cite{barbieri2020tweeteval}, a widely adopted benchmark for social media sentiment analysis, producing continuous scores from -1 (negative) to +1 (positive). 

\textbf{Framing Classification: } We classify each Reddit post into one of five framing categories (THREAT, ECONOMIC, NEUTRAL, HUMANITARIAN, DIPLOMACY) using GPT-4o-mini (API version: July 2024, temperature = 0, seed = 42 for reproducibility) within a ``Codebook LLM'' framework, which treats the model as a clear-cut measurement instrument rather than a black box. The prompt provides category definitions and requires a single label output; the model also returns a confidence score.
Detailed prompt instructions and category definitions are provided in Appendix~\ref{app:llm_prompt}. To ensure classification quality, we retained only predictions with confidence $\geq 0.90$, yielding 27,863 posts (94\% of the original sample). For DiD analysis, we convert categorical framing to a continuous \textit{diplomacy scale}: THREAT = $-2$, DIPLOMACY = $+2$, and NEUTRAL/ECONOMIC/HUMANITARIAN = $0$. This scale captures the primary distinction between military threat framing and diplomatic engagement framing, allowing direct quantification of how discourse shifts between adversarial and cooperative orientations. While economic sanctions can function as coercive instruments in international relations, we treat ECONOMIC framing as neutral in our primary specification because our theoretical focus is on the rhetorical distinction between \textit{military threats} and \textit{diplomatic dialogue}---the two poles most directly affected by summit diplomacy. A robustness check treating ECONOMIC as partially coercive (ECONOMIC = $-1$) yields similar or stronger results, suggesting our main specification is conservative (see Appendix~\ref{app:robustness}). We aggregate individual post scores to monthly averages, yielding a continuous outcome variable for each country-month observation. This operationalization follows prior work emphasizing the distinction between affective tone and substantive narrative positioning.

\textbf{Human Annotation Benchmark: } To validate our LLM-based classification in a domain-sensitive setting, we constructed a gold-standard human annotation benchmark. Two military officers with expertise in North Korean affairs independently labeled 500 stratified samples (4 countries $\times$ 3 periods $\times$ 5 frames). We provide detailed information in the Results section.

\subsection{Causal Identification}
We estimate the causal impact of summit diplomacy using a Difference-in-Difference (DiD) design, a widely used quasi-experimental approach for evaluating the effects of discrete political events in observational data~\cite{berger2020tarp,yang2022effects}. Following Kumarswamy et al.~\cite{kumarswamy2025}, who used a control platform to isolate policy effects, we leverage multiple control countries to separate summit impacts from global geopolitical confounds.
By using China, Iran, and Russia as counterfactuals, we control for external shocks such as the U.S.-China trade war, JCPOA withdrawal tensions, and the Mueller investigation, respectively. This design ensures that observed changes in North Korea discourse are attributable to summit diplomacy rather than general fluctuations in foreign policy attention or sentiment.

\textbf{Difference-in-Difference Specification:} Our main specification uses two-way fixed effects:
\begin{equation}
Y_{g,t} = \alpha_g + \gamma_t + \beta \big(\text{Treat}_g \times \text{Post}_t\big) + \epsilon_{g,t},
\end{equation}
where $Y_{g,t}$ is the monthly average sentiment (or framing) for group $g$ in month $t$, $\alpha_g$ and $\gamma_t$ denote group and month fixed effects, and $\beta$ captures the DiD estimate of the summit effect. We report heteroskedasticity-robust standard errors clustered at the group level. We explicitly verified the parallel trends assumption, as detailed in the Data Collection subsection above.

\textbf{Causal Scope:}
Our primary causal identification is defined at the post level, where the unit of analysis is the agenda-setting initial post. We additionally estimate DiD at the comment level to assess whether post-level effects extend to highly visible audience responses. These comment-level estimates are interpreted as supporting evidence of propagation and external validity, rather than as independent causal identification.

\subsection{Network Structure Analysis}
Beyond content-level changes, we examine whether diplomatic events reorganize the \textit{structure} of discourse. We adopt a GraphRAG-inspired indexing pipeline~\cite{edge2024graphrag}, leveraging its entity and relationship extraction capabilities to construct structured knowledge graphs. Recent advances have demonstrated that Large Language Models (LLMs) can effectively extract structured knowledge from unstructured text across complex domains, often outperforming traditional information extraction methods~\cite{Anuyah2025CoDeKG,Yang2025SepsisKG}. This approach has proven particularly valuable for analyzing political discourse, where LLMs can identify key actors, events, and their relationships from noisy media content~\cite{Arslan2024PoliticalRAG,Fadda2025LLM_KG_Viewpoints}. While GraphRAG is primarily designed for retrieval-augmented generation, we utilize only its initial \textbf{indexing phase} (which extracts entities and relationships via LLM to build a structured knowledge graph) as a scalable, automated method for constructing the inputs needed for Discourse Network Analysis (DNA)~\cite{leifeld2016policy}. This methodology aligns with recent frameworks that use LLM-constructed graphs to capture complex multi-level relationships and unification of fragmented evidence~\cite{Feng2025BioRAG,Ling2026LLM_KG_Review}.
Unlike simple keyword co-occurrence networks, which cannot distinguish whether two entities co-occur in a ``threat'' or ``negotiation'' context, this approach extracts semantically meaningful relationships, enabling analysis of \textit{narrative connectivity} rather than mere lexical proximity. We utilize this pipeline to generate knowledge graphs from Reddit discussions for each analysis period.

\textbf{Graph Construction and Customization: }We customized the graph extraction prompts to capture domain-specific constraints. We extended the default entity taxonomy to include \textbf{WEAPON} (e.g., ``ICBM,'' ``nuclear warhead'') and \textbf{POLICY} (e.g., ``denuclearization,'' ``maximum pressure'') categories, ensuring the retrieval of security-relevant concepts central to North Korea discourse.

\textbf{Hierarchical Framing Classification: }To ensure methodological consistency between content and structural analyses, we applied the same ``Codebook LLM'' classifier (GPT-4o-mini) used for individual posts to the structural outputs of the graph-based indexing pipeline:
\begin{enumerate}
    \item \textbf{Edge Classification}: We classified the extracted relationship descriptions between entities to determine if the structural link represented a ``Threat,'' ``Diplomacy,'' or other connection type.
    \item \textbf{Community Classification}: We classified the generated summaries of each community to identify the dominant narrative orientation of that cluster.
\end{enumerate}

\textbf{Network Metrics: } We compute three key structural metrics to test for discourse reorganization:
\begin{enumerate}
    \item \textbf{Network Density}: Measures structural consolidation and discourse cohesion (testing for fragmentation vs. unification).
    \item \textbf{Edge Framing Distribution}: Quantifies the proportion of adversarial vs. diplomatic relationships, providing a structural test of asymmetric persistence in framing.
    \item \textbf{Community Composition}: Identifies thematic diversification by tracking the emergence of non-security clusters (e.g., humanitarian, economic) in the community structure.
\end{enumerate}
This structural analysis tests whether content-level framing shifts are accompanied by reorganization of discourse networks.

%% file: sections/results.tex
\section{Results}
\label{res}

\subsection{Content Changes in Online Discourse (RQ1a)}
To explicitly compare the relative sensitivity of these measures, we report their period-to-period changes.
From P1 to P2, the Diplomacy/Threat framing ratio increased by 204\%, whereas average sentiment increased more modestly in magnitude (Table~\ref{tab:did_sentiment_summary}).
Following the Hanoi failure, both measures declined, reinforcing the view that distinguishing framing from sentiment is critical for capturing asymmetric persistence.

\begin{figure*}[t]
\centering
    \subfloat[Singapore Summit Effect (P1 \textrightarrow\ P2)]{\includegraphics[width=0.38\textwidth]{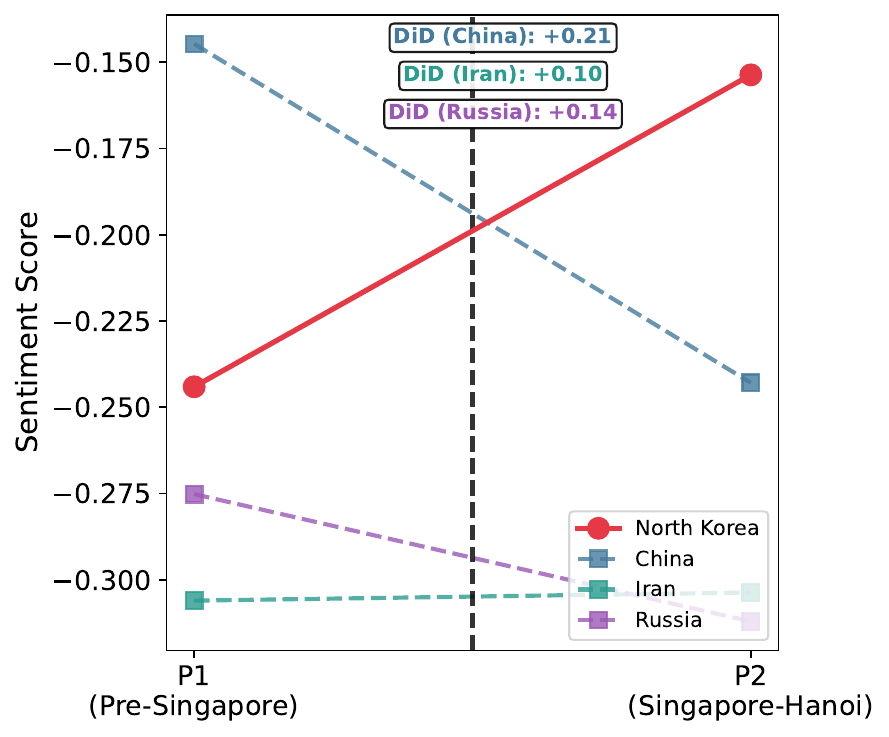}\label{fig:sentiment_did_a}}
    \subfloat[Hanoi Summit Effect (P2 \textrightarrow\ P3)]{\includegraphics[width=0.38\textwidth]{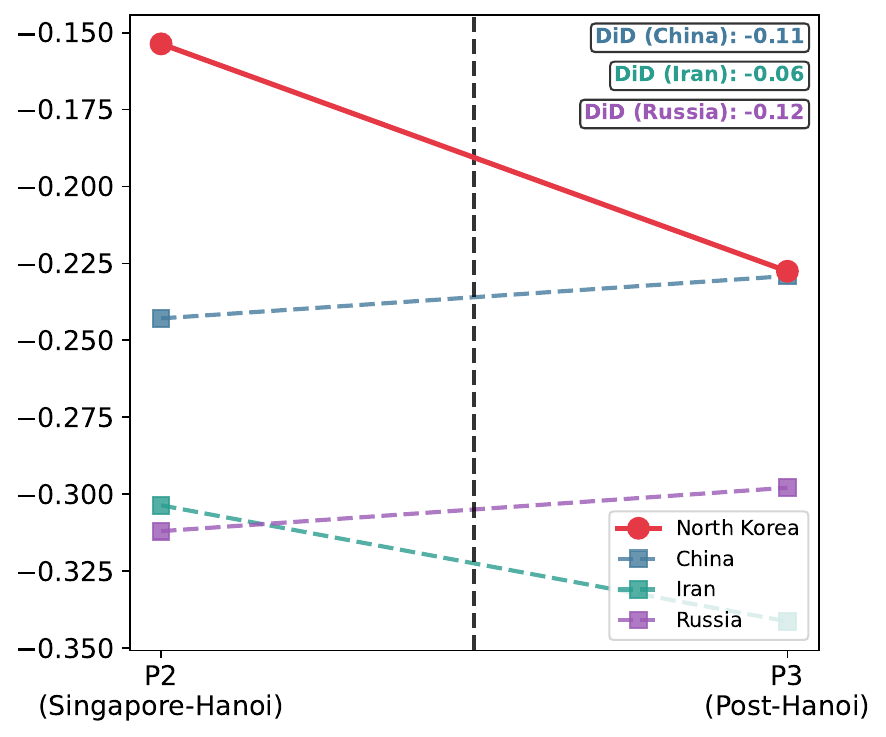}\label{fig:sentiment_did_b}}
\caption{Sentiment Difference-in-Differences Visualization. Pre-treatment trends (P1) are visually parallel across groups, supporting the validity of the DiD design. The bold black dashed line marks the intervention point (summit date).}
\label{fig:sentiment_did}
\end{figure*}

\textbf{Impact on Sentiment: } Figure~\ref{fig:sentiment_did} visualizes the DiD estimates. The Singapore Summit (P1$\rightarrow$P2) produced a statistically significant positive shift in sentiment toward North Korea across all control groups (Table~\ref{tab:did_sentiment_summary}), with causal estimates ranging from +0.10 to +0.21 ($p < 0.001$). In contrast, the failure of the Hanoi Summit (P2$\rightarrow$P3) resulted in a significant negative shift in sentiment, ranging from -0.06 to -0.12 ($p < 0.001$). 
While these sentiment shifts are statistically significant, sentiment alone cannot reliably distinguish between threat-oriented and diplomacy-oriented discourse. Both exhibit predominantly negative valence with substantial overlap in their distributions (THREAT: $M = -0.27$, $SD = 0.27$; DIPLOMACY: $M = -0.08$, $SD = 0.34$). Consequently, an improvement in sentiment may reflect either a genuine shift toward diplomatic framing or merely a softening of tone within the same rhetorical frame. This limitation underscores why framing analysis is essential to capture substantive rhetorical changes.

\begin{table}[t]
\centering
\caption{Sentiment Difference-in-Differences Results}
\resizebox{0.75\linewidth}{!}{
\begin{tabular}{llcc}
\toprule
\textbf{Event} & \textbf{Control} & \textbf{DiD Est.} & \textbf{95\% CI} \\
\midrule
Singapore & China & +0.21*** & [0.18, 0.24] \\
(P1$\rightarrow$P2) & Iran & +0.10*** & [0.06, 0.13] \\
 & Russia & +0.14*** & [0.11, 0.16] \\
\midrule
Hanoi & China & -0.11*** & [-0.14, -0.08] \\
(P2$\rightarrow$P3) & Iran & -0.06** & [-0.10, -0.02] \\
 & Russia & -0.12*** & [-0.15, -0.10] \\
\bottomrule
\multicolumn{4}{l}{\footnotesize *$p<0.05$, **$p<0.01$, ***$p<0.001$} \\
\end{tabular}
}
\label{tab:did_sentiment_summary}
\end{table}

\begin{figure*}[t]
\centering
    \subfloat[Singapore Summit Effect (P1 \textrightarrow\ P2)]{\includegraphics[width=0.38\textwidth]{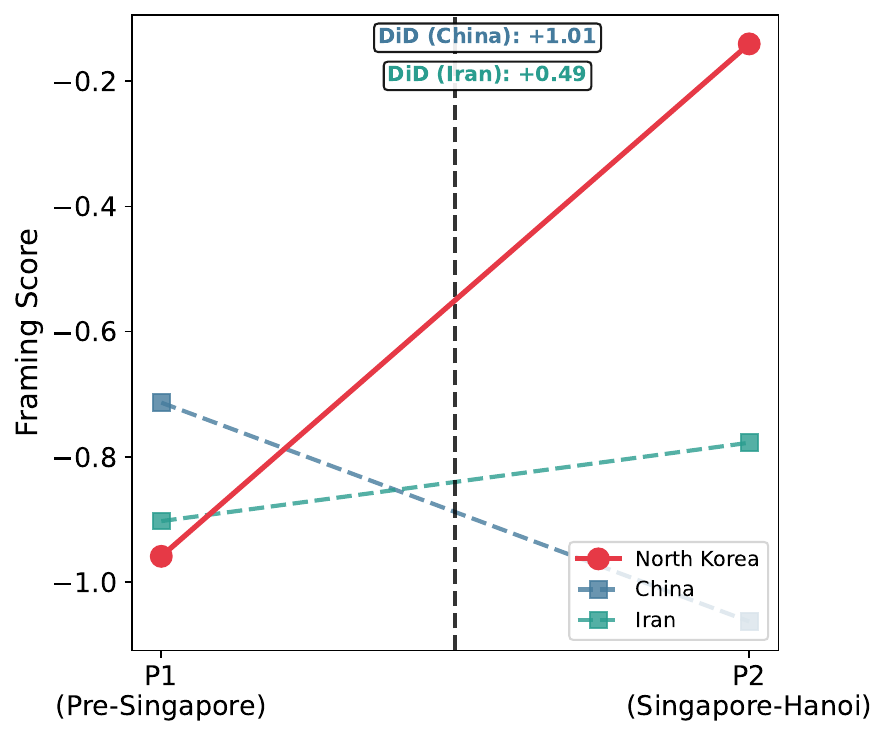}\label{fig:did_a}}
    \subfloat[Hanoi Summit Effect (P2 \textrightarrow\ P3)]{\includegraphics[width=0.38\textwidth]{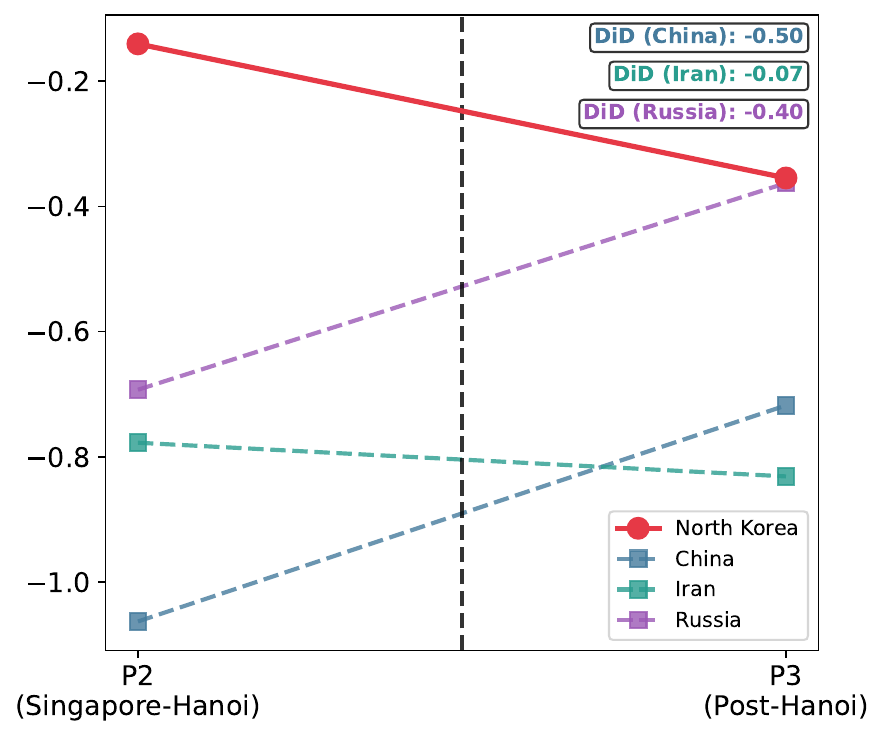}\label{fig:did_b}}
\caption{Framing Difference-in-Differences Visualization. Pre-treatment trends are visually parallel for China and Iran, validating their use as control groups. Russia is excluded from (A) due to parallel trends violation (p=0.01) but included in (B) where parallel trends are satisfied. The bold black dashed line marks the intervention point (summit date).}
\label{fig:did}
\end{figure*}

\textbf{Impact on Framing:} We next examine changes in framing, where higher values indicate diplomacy-oriented narratives and lower values indicate threat-oriented narratives. 
Figure~\ref{fig:did} visualizes the DiD estimates.

\begin{table}[t]
\centering
\caption{Framing Difference-in-Differences Results}
\resizebox{0.75\linewidth}{!}{
\begin{tabular}{llcc}
\toprule
\textbf{Event} & \textbf{Control} & \textbf{DiD Est.} & \textbf{95\% CI} \\
\midrule
Singapore & China & +1.17*** & [0.64, 1.69] \\
(P1$\rightarrow$P2) & Iran & +0.69* & [0.17, 1.22] \\
\midrule
Hanoi & China & -0.56* & [-1.09, -0.03] \\
(P2$\rightarrow$P3) & Iran & -0.16 & [-0.72, 0.40] \\
& Russia & -0.54 & [-1.15, 0.06] \\
\bottomrule
\multicolumn{4}{l}{\footnotesize * $p<0.05$, ** $p<0.01$, *** $p<0.001$} \\
\end{tabular}
\label{tab:did_framing_summary}
}
\end{table}

The Singapore Summit led to a pronounced shift toward diplomacy-oriented framing
(Table~\ref{tab:did_framing_summary}), with DiD estimates ranging from +0.69 to +1.17
across valid control groups. Following the collapse of the Hanoi Summit, framing
partially shifted back toward threat-oriented narratives. The China control group
exhibits a modest and marginal reversal ($-0.56$, $p = 0.048$), while no statistically
detectable reversion is observed for the remaining control groups (Iran, Russia).
This asymmetry---a strong and consistent shift following diplomatic success but only
limited and non-systematic reversal following diplomatic failure---is consistent with
the asymmetric persistence (ratchet) pattern examined below. Notably, the magnitude
of framing shifts substantially exceeds that of sentiment changes, indicating that
diplomatic events more strongly reshape how North Korea is framed than how it is
emotionally evaluated in online discourse.

To verify that these findings are not artifacts of the diplomacy scale construction
(THREAT = $-2$, DIPLOMACY = $+2$), we conducted robustness checks using binary outcome
indicators. DiD analyses of the proportion of THREAT-framed posts reveal substantial
declines following the Singapore Summit across control groups ($-26.9$ pp with China,
$p = 0.004$; $-17.0$ pp with Iran, $p = 0.073$; $-29.1$ pp with Russia, $p = 0.001$).
In contrast, increases in DIPLOMACY framing are more uneven, reaching statistical
significance primarily for China ($+21.0$ pp, $p < 0.001$). Taken together, these
results suggest that summit diplomacy primarily operated by reducing threat-oriented
framing rather than generating symmetric increases in explicit diplomatic endorsement,
reinforcing the asymmetric persistence observed in the main analyses.

\subsection{Asymmetric Effect Persistence (RQ1b)} 
To test whether diplomatic effects exhibit asymmetric persistence (the ratchet effect), we compare the magnitude of framing shifts following the Singapore Summit ($|P1 \rightarrow P2|$) against subsequent reversals following the Hanoi failure ($|P2 \rightarrow P3|$). If the asymmetric persistence effect holds, the reversal effect should be significantly smaller than the original effect.
Table~\ref{tab:ratchet_validation} presents bootstrap validation results across multiple dimensions. For framing dimensions, we use a combined score (DIPLOMACY\% $-$ THREAT\%) to capture the overall shift from threat-dominant to diplomacy-dominant discourse. Ratios significantly below 1.0 indicate incomplete reversal.

\begin{table}[t]
\centering
\small
\setlength{\tabcolsep}{3pt}
\caption{Asymmetric Persistence Validation: Unlike sentiment, framing shifts exhibit asymmetric persistence (Ratio $\ll$ 1.0). Content, Edge, and Community refer to framing metrics at the post, relationship, and cluster levels, respectively.}
\resizebox{\linewidth}{!}{
\begin{tabular}{lcccc}
\toprule
\textbf{Metric} & \textbf{$\Delta$(P1$\to$P2)} & \textbf{$\Delta$(P2$\to$P3)} & \textbf{Ratio} & \textbf{95\% CI} \\
\midrule
Content & 28.9pp & 11.4pp & 0.39* & [0.15, 0.65] \\
Edge & 38.6pp & 4.3pp & 0.11* & [0.01, 0.24] \\
Community & 37.2pp & 1.4pp & 0.24* & [0.01, 0.72] \\
Sentiment & 0.095 & 0.109 & 1.15 & [0.97, 1.35] \\
\bottomrule
\multicolumn{5}{l}{\footnotesize pp = percentage points; * 95\% CI excludes 1.0 (significant asymmetry)} \\
\end{tabular}
}
\label{tab:ratchet_validation}
\end{table}

The results are consistent with the asymmetric persistence effect for framing dimensions. Content framing shows a reversal ratio of 0.39 (95\% CI: [0.15, 0.65]), indicating that the Hanoi failure reversed only 39\% of the Singapore Summit's framing effect. Edge-level framing exhibits an even stronger ratchet-like pattern (ratio = 0.11), and community framing shows similar persistence (ratio = 0.24). Critically, all framing dimensions have 95\% CIs excluding 1.0. In contrast, sentiment shows a ratio of 1.15 (95\% CI including 1.0), indicating that emotional valence fully reverted following diplomatic failure.
The DiD analysis provides additional statistical evidence for this asymmetry. During the Singapore Summit period (P1$\rightarrow$P2), framing shifts were statistically significant across all valid control groups (China: $p < 0.001$; Iran: $p = 0.014$). In contrast, during the Hanoi reversal period (P2$\rightarrow$P3), only one control group (China) showed a marginally significant effect ($p = 0.048$), while Iran ($p = 0.58$) and Russia ($p = 0.087$) showed no significant reversal. This pattern---where diplomatic engagement produces robust, statistically significant shifts but diplomatic failure produces weaker, less consistent reversals---provides evidence consistent with such a dynamic.

\textbf{Summary: }This divergence between framing and sentiment provides important theoretical insight (RQ1a): while affective responses to diplomacy may be transient, interpretive frames—how North Korea is discussed as a threat versus a negotiation partner—exhibit greater persistence. The asymmetric persistence (RQ1b) appears to be specific to framing rather than sentiment, suggesting that diplomatic engagement primarily reshapes cognitive interpretive schemas rather than emotional evaluations.

\subsection{Structural Reorganization of Discourse Networks (RQ2)}
To assess whether observed framing changes are reflected in the organization of discourse itself, we analyze discourse networks constructed from Reddit discussions across the three diplomatic phases. Using a graph-based indexing approach~\cite{edge2024graphrag}, we extract entities and relationships from 10,659 documents, constructing knowledge graphs that capture how actors, events, and concepts are interconnected in public discourse.

\textbf{Network Topology Changes:} Table~\ref{tab:network_topology} presents network-level metrics across the three periods. Despite substantial reductions in nodes and edges from P1 to P3, network density increased markedly (+146\% from P1 to P2), signifying \textit{structural consolidation}: discussion became highly interconnected around core actors (e.g., ``Trump,'' ``Kim,'' ``summit''). The proportion of nodes in the largest connected component increased from 78\% (P1) to 91\% (P2), indicating that diplomatic engagement integrated previously fragmented discourse threads.

\input{tables/network_topology}

\textbf{Relationship Framing Shifts: }
Table~\ref{tab:relationship_framing} reveals a pronounced structural inversion. In P1, threat-oriented relationships dominated (48.4\%), but following Singapore (P2), threat framing dropped to 28.0\% ($-20.4$ pp) while diplomacy framing nearly doubled ($+18.3$ pp). 
\input{tables/relationship_framing}
Critically, the Hanoi failure (P3) triggered only partial reversion: threat framing remained low (25.6\%) and diplomacy framing decreased modestly ($-6.7$ pp). This asymmetric pattern suggests that once diplomatic connections are established in the public's conceptual map, the underlying relationship structure resists reverting to a pure threat footing.

\textbf{Community Theme Evolution: }
Community structures mirrored edge-level shifts (Table~\ref{tab:community_frame_orientation}). In P2, diplomacy-oriented communities surged ($+17.9$ pp) while threat communities declined ($-19.3$ pp).
\input{tables/community_frame_orientation}
In P3, rather than reverting to threat dominance, the discourse structure diversified: threat communities continued declining (26.2\%), while humanitarian ($+9.4$ pp) and economic ($+5.1$ pp) communities expanded, suggesting a broadening of the discursive agenda.

\textbf{Summary: }These analyses show that the Singapore Summit not only shifted edge-level framing (Table~\ref{tab:relationship_framing}) but also reorganized community-level thematic structure. The decline in threat-oriented framing and rise of diplomacy-oriented framing from P1 to P2, coupled with the subsequent diversification in P3, suggests that diplomatic events create cascading effects across multiple levels of discourse organization.

\subsection{Audience Response Analysis (RQ3)}
\label{subsec:audience_response}

We next examine whether the post-level effects identified in RQ1 and RQ2
are reflected in highly visible audience responses.
Our goal is not to treat comments as independent causal units,
but to assess whether framing shifts associated with diplomatic events
extend beyond agenda-setting posts into participatory discourse.

\textbf{Content-Level Audience Responses: }
We first analyze sentiment and framing patterns in audience comments.
Difference-in-Difference (DiD) analyses on highly visible comments
($N=255{,}391$; Table~\ref{tab:dataset}) reveal significant sentiment improvements
during the Singapore Summit period (Table~\ref{tab:comment_sentiment_did}).
Following the Hanoi failure, however, audience sentiment largely reverts
toward pre-summit levels, mirroring the transient pattern observed at the post level.

\begin{table}[t]
\centering
\caption{Comment Sentiment Difference-in-Differences Results}
\resizebox{0.9\linewidth}{!}{
\begin{tabular}{llcc}
\toprule
\textbf{Event} & \textbf{Control} & \textbf{DiD Est.} & \textbf{95\% CI} \\
\midrule
Singapore & China & +0.04* & [0.01, 0.08] \\
(P1$\rightarrow$P2) & Iran & +0.04 & [-0.00, 0.09] \\
 & Russia & +0.05** & [0.02, 0.09] \\
\midrule
Hanoi & China & -0.05 & [-0.18, 0.08] \\
(P2$\rightarrow$P3) & Iran & -0.00 & [-0.04, 0.04] \\
 & Russia & -0.03* & [-0.05, -0.00] \\
\bottomrule
\multicolumn{4}{l}{\footnotesize *$p<0.05$, **$p<0.01$, ***$p<0.001$} \\
\end{tabular}
}
\label{tab:comment_sentiment_did}
\end{table}

This contrast underscores a key distinction between affective responses
and narrative framing in audience discourse.
In particular, while sentiment is fleeting, structural framing persists—a pattern confirmed by calculating the magnitude of reversal following the Hanoi collapse relative to the initial summit shift (Table~\ref{tab:comment_ratchet_validation}).
\begin{table}[t]
\centering
\small
\setlength{\tabcolsep}{3pt}
\caption{Comment Ratchet Effect Validation: Unlike framing (which ratchets), sentiment exhibits full reversion (Ratio $>$ 1.0) when measured by monthly aggregation.}
\resizebox{0.95\linewidth}{!}{
\begin{tabular}{lcccc}
\toprule
\textbf{Metric} & \textbf{$\Delta$(P1$\to$P2)} & \textbf{$\Delta$(P2$\to$P3)} & \textbf{Ratio} & \textbf{95\% CI} \\
\midrule
Content & 11.6pp & 2.1pp & 0.18* & [0.08, 0.26] \\
Edge & 23.7pp & 9.5pp & 0.40* & [0.31, 0.48] \\
Community & 22.1pp & 11.9pp & 0.54* & [0.17, 0.83] \\
Sentiment & 0.046 & 0.074 & 1.63 & [0.06, 8.85] \\
\bottomrule
\multicolumn{5}{l}{\footnotesize pp = percentage points; * Significant asymmetry (Ratio $<$ 1.0)} \\
\end{tabular}
}
\label{tab:comment_ratchet_validation}
\end{table}
In contrast, framing patterns in comments closely mirror post-level dynamics.
Threat-oriented discourse collapses during the summit period
and does not statistically revert to pre-summit levels following the diplomatic failure
(Table~\ref{tab:comment_framing_did}).

\begin{table}[t]
\centering
\caption{Comment Framing Difference-in-Differences Results}
\resizebox{0.9\linewidth}{!}{
\begin{tabular}{llcc}
\toprule
\textbf{Event} & \textbf{Control} & \textbf{DiD Est.} & \textbf{95\% CI} \\
\midrule
Singapore & China & +0.48** & [0.14, 0.82] \\
(P1$\rightarrow$P2) & Iran & +0.44* & [0.06, 0.82] \\
 & Russia & +0.32 & [-0.03, 0.67] \\
\midrule
Hanoi & China & -0.15 & [-0.63, 0.33] \\
(P2$\rightarrow$P3) & Iran & +0.07 & [-0.54, 0.67] \\
 & Russia & -0.18 & [-0.49, 0.12] \\
\bottomrule
\multicolumn{4}{l}{\footnotesize *$p<0.05$, **$p<0.01$, ***$p<0.001$} \\
\end{tabular}
}
\label{tab:comment_framing_did}
\end{table}

The persistence of diplomacy-oriented framing in comments,
alongside the transience of sentiment,
reinforces the asymmetric pattern observed at the post level
and suggests continuity in how diplomatic engagement reshapes narrative orientation.

\textbf{Structural-Level Audience Responses: }
We next assess whether these content-level patterns are reflected
in the structure of audience discourse networks. Table~\ref{tab:app_comment_structure} shows the graph-based analyses of comment networks, indicating that the post-Hanoi period is characterized not by a return to purely confrontational structures,
but by a mixed configuration in which diplomatic connections remain salient
alongside re-emerging threats.
In particular, diplomacy-oriented edges remain elevated relative to pre-summit baselines,
while humanitarian and sanctions-related themes expand.
Together, these patterns point to a reconfiguration of audience discourse
toward sustained, if contested, engagement rather than full reversion.

\begin{table}[h]
\centering
\caption{Structural Framing Distribution in Comment Networks}
\resizebox{0.85\linewidth}{!}{
\begin{tabular}{lccc}
\toprule
\textbf{Frame} & \textbf{P1} & \textbf{P2} & \textbf{P3} \\
\midrule
\multicolumn{4}{l}{\textit{Community-Level}} \\
THREAT & 18.6\% & 11.3\% & 12.1\% \\
DIPLOMACY & 13.8\% & \textbf{28.6\%} & 17.5\% \\
HUMANITARIAN & 5.4\% & 4.8\% & \textbf{10.8\%} \\
NEUTRAL & 60.2\% & 51.9\% & 58.0\% \\
Count ($N$) & 746 & 231 & 371 \\
\midrule
\multicolumn{4}{l}{\textit{Edge-Level}} \\

THREAT & 29.1\% & 12.2\% & 18.6\% \\
DIPLOMACY & 15.9\% & \textbf{22.0\%} & 19.7\% \\
HUMANITARIAN & 3.8\% & 1.9\% & \textbf{6.5\%} \\
NEUTRAL & 45.1\% & 58.3\% & 48.5\% \\
Count ($N$) & 10,730 & 3,249 & 4,522 \\
\bottomrule
\end{tabular}
}
\label{tab:app_comment_structure}
\end{table}

\textbf{Summary: }
Taken together, audience-level analyses reinforce the post-level conclusions
in two ways.
First, framing shifts associated with summit diplomacy are preserved
in highly visible audience interpretations, indicating that these effects
are not confined to agenda-setting posts.
Second, structural patterns in comment networks parallel the reorganization
observed at the post level, supporting a system-level interpretation
of discourse change.
Importantly, these results do not constitute independent causal estimates;
rather, they provide convergent evidence that post-level effects
are reflected in participatory discourse.

\subsection{Validation of LLM-Based Framing (RQ4)}
To evaluate whether LLM-based framing classification can approximate expert human judgment in geopolitical discourse analysis, we constructed a gold-standard benchmark of 500 Reddit posts using stratified sampling across countries (North Korea, China, Iran, Russia), time periods (P1--P3), and frames (THREAT, ECONOMIC, NEUTRAL, HUMANITARIAN, DIPLOMACY). Two domain experts (military officers with expertise in North Korea and international security) independently annotated all samples following an iterative codebook refinement process.

\textbf{Inter-Rater Reliability:} We first assess annotation reliability on the independently produced labels prior to consensus. Overall agreement between annotators was 81.8\%, with Cohen's $\kappa$ = 0.75, indicating substantial agreement.
Reliability was consistent across all framing categories, ranging from 79.6\% (DIPLOMACY) to 83.9\% (HUMANITARIAN), suggesting that the codebook definitions enabled uniform annotation quality across diverse framing contexts.

\textbf{LLM--Human Agreement:} Using the final consensus labels as ground truth, we compare LLM predictions against the benchmark annotations. Overall accuracy was 85.0\%, with Cohen's $\kappa$ = 0.79, indicating substantial agreement. Performance varied by framing category: the model achieved strong performance on all frames (F1 $\geq$ 0.75), with the highest performance on THREAT (F1 = 0.88) and NEUTRAL (F1 = 0.88). We report the full per-class precision/recall/F1 in Table~\ref{tab:llm_human_metrics}.

\begin{table}[t]
\centering
\small
\setlength{\tabcolsep}{4pt}
\caption{LLM vs.\ Human (Consensus) Validation Metrics}
\resizebox{0.9\linewidth}{!}{
\begin{tabular}{lcccc}
\toprule
\textbf{Frame} & \textbf{Prec.} & \textbf{Rec.} & \textbf{F1} & \textbf{Support} \\
\midrule
THREAT & 0.91 & 0.86 & 0.88 & 97 \\
ECONOMIC & 0.78 & 0.81 & 0.79 & 52 \\
NEUTRAL & 0.85 & 0.90 & 0.88 & 225 \\
HUMANITARIAN & 0.70 & 0.81 & 0.75 & 32 \\
DIPLOMACY & 0.88 & 0.77 & 0.82 & 108 \\
\midrule
\textbf{Overall} &  &  & \textbf{Acc. 0.85} &  \\
\textbf{Macro Avg.} & 0.83 & 0.83 & \textbf{0.83} &  \\
\bottomrule
\end{tabular}
}
\label{tab:llm_human_metrics}
\end{table}

\textbf{Error Analysis and Implications:} Qualitative inspection of disagreements suggests that most LLM errors arise in edge cases where multiple frames co-occur, and the primary emphasis is ambiguous. Specifically, the LLM over-applied the ``factual reporting = NEUTRAL'' heuristic to diplomatic event coverage (17 cases), classifying summits and negotiations as neutral when human annotators labeled them as DIPLOMACY. Similarly, the ``individual harm = HUMANITARIAN'' rule was triggered by posts merely mentioning individuals (8 cases), even when the overall framing was neutral. Importantly, these errors are unlikely to systematically favor one diplomatic phase over another because the benchmark was stratified across periods and countries, with error rates ranging from 11.6\% (Iran) to 18.0\% (North Korea). We provide detailed qualitative examples of these classification disagreements in Appendix~\ref{app:error_analysis}.

\textbf{Summary: }These results support the validity of using LLM-based framing classification for large-scale causal analysis.

%% file: tables/network_topology.tex
\begin{table}[t]
\centering
\setlength{\tabcolsep}{3pt}
\caption{Network Topology Metrics Across Periods}
\small
\resizebox{0.9\linewidth}{!}{
\begin{tabular}{lccccc}
\toprule
\textbf{Metric} & \textbf{P1} & \textbf{P2} & \textbf{P3} & \textbf{$\Delta_1$} & \textbf{$\Delta_2$} \\
\midrule
Nodes & 2,656 & 1,043 & 879 & $-61\%$ & $-16\%$ \\
Edges & 4,552 & 1,726 & 1,429 & $-62\%$ & $-17\%$ \\
Density ($\times10^{-3}$) & 1.3 & 3.2 & 3.7 & $+146\%$ & $+16\%$ \\
Avg Degree & 3.43 & 3.31 & 3.25 & $-3\%$ & $-2\%$ \\
Clustering & 0.215 & 0.198 & 0.210 & $-8\%$ & $+6\%$ \\
Components & 25 & 23 & 16 & $-8\%$ & $-30\%$ \\
\bottomrule
\end{tabular}
}
\label{tab:network_topology}

\footnotesize{$\Delta_1$: P1$\rightarrow$P2; $\Delta_2$: P2$\rightarrow$P3}
\end{table}

%% file: tables/relationship_framing.tex
\begin{table}[t]
\centering
\caption{Relationship Framing Distribution Across Periods}
\resizebox{0.75\linewidth}{!}{
\begin{tabular}{lccc}
\toprule
\textbf{Frame} & \textbf{P1} & \textbf{P2} & \textbf{P3} \\
\midrule
THREAT & 48.4\% & 28.0\% & 25.6\% \\
DIPLOMACY & 20.6\% & 38.9\% & 32.2\% \\
NEUTRAL & 22.0\% & 24.6\% & 26.8\% \\
ECONOMIC & 5.7\% & 6.2\% & 8.3\% \\
HUMANITARIAN & 3.3\% & 2.3\% & 7.1\% \\
\midrule
\textit{Total edges} & 4,838 & 1,838 & 1,530 \\
\bottomrule
\end{tabular}
}
\label{tab:relationship_framing}
\end{table}

%% file: tables/community_frame_orientation.tex
\begin{table}[t]
\centering
\caption{Community Frame Orientation by Period}
\resizebox{0.75\linewidth}{!}{
\begin{tabular}{lccc}
\toprule
\textbf{Dominant Frame} & \textbf{P1} & \textbf{P2} & \textbf{P3} \\
\midrule
THREAT & 52.0\% & 32.7\% & 26.2\% \\
DIPLOMACY & 24.6\% & 42.5\% & 34.6\% \\
NEUTRAL & 13.0\% & 17.0\% & 16.8\% \\
ECONOMIC & 3.1\% & 3.3\% & 8.4\% \\
HUMANITARIAN & 7.3\% & 4.6\% & 14.0\% \\
\midrule
\textit{Total communities} & 354 & 153 & 107 \\
\bottomrule
\end{tabular}
}
\label{tab:community_frame_orientation}
\end{table}

%% file: sections/discussion.tex
\section{Discussion}
\textbf{Asymmetric Persistence in Diplomatic Framing: } Our findings indicate an asymmetry in how diplomatic events shape online discourse. The 2018 Singapore Summit produced substantial shifts toward diplomacy-oriented framing across multiple levels of analysis, including individual posts (+0.85 to +1.28 in DiD estimates), relational structures (THREAT edges: 48.4\% $\rightarrow$ 28.0\%), and community organization (THREAT communities: 52.0\% $\rightarrow$ 32.7\%). In contrast, the subsequent failure of the 2019 Hanoi Summit triggered only partial reversals. Rather than returning to pre-summit baselines, discourse stabilized at an intermediate state between threat-dominant and diplomacy-oriented narratives.
We refer to this pattern as \textbf{asymmetric persistence}, using the term ``ratchet effect'' as a descriptive metaphor. The term is used solely as an illustrative metaphor and does not denote a formal causal mechanism or theoretical model. Across content-level framing (reversal ratio = 0.43), edge-level relationships (0.11), and community-level orientations (0.24), the magnitude of reversal following diplomatic failure was consistently smaller than the magnitude of the initial shift following diplomatic success. Notably, this pattern does not hold for sentiment, which fully reverted after the Hanoi Summit. This divergence suggests that diplomatic events reshape \textit{how} adversaries are discussed more durably than \textit{how they are emotionally evaluated}. Rather than constituting independent causal estimates, comment-level analyses illustrate how post-level effects extend into participatory discourse. This leads to the observation that diplomatic events can restructure not only agenda-setting content but also the interpretive dynamics of online publics.

One plausible explanation for this asymmetry lies in the structural reorganization of discourse. Our network analysis indicates that diplomatic engagement consolidates discussion around a smaller set of central actors and themes, increasing narrative connectivity and reducing fragmentation. Once such connections are established (linking adversaries to negotiation, dialogue, and institutional processes), they appear to persist even when subsequent events undermine diplomatic optimism. In this sense, diplomatic success may introduce new interpretive linkages that are not immediately undone by later failure.
Our results suggest that framing responds to positive diplomatic engagement in a more \textit{structurally persistent} manner than sentiment. This asymmetry highlights the value of distinguishing between affective reactions and interpretive structures when evaluating the public impact of diplomacy. While sentiment captures short-lived emotional responses to events, framing reflects deeper narrative orientations that can endure beyond immediate political outcomes~\cite{bestvater2023sentiment}. This divergence is further reinforced by comment-level patterns: even when affective responses among audiences remain elastic, the relational and thematic structure of discourse continues to privilege diplomatic interpretations over threat-oriented ones.
Alternative mechanisms could potentially explain the observed asymmetric persistence. \textit{Agenda saturation}, for example, might suggest that diplomatic themes simply became part of the general news cycle, making reversal less likely due to diminished novelty. \textit{Topic fatigue} could imply that users lost interest in adversarial framing altogether, rather than genuinely shifting their interpretive schemas. A third alternative, \textit{background knowledge absorption}, posits that diplomatic frames became embedded as default contextual knowledge rather than actively maintained discourse positions. However, several features of our findings distinguish a structural ratchet interpretation from these alternatives. First, if agenda saturation or topic fatigue were driving, we would expect \textit{uniform} decline across all framing categories---yet THREAT framing specifically decreased while DIPLOMACY framing increased. Second, the network-level reorganization (increased centralization, reduced fragmentation) indicates active structural change rather than passive attention decay. Third, the differential behavior of sentiment versus framing is inconsistent with general fatigue explanations: sentiment fully reverted while framing persisted, suggesting that interpretive frames---rather than attention levels---are the locus of durability.

\textbf{Multi-Level Discourse Reorganization:} Beyond content-level shifts, our graph-based network analysis reveals that diplomatic events reorganize discourse at multiple levels~\cite{vandenhole2025discourse}. The Singapore Summit produced coordinated shifts: threat edges dropped 20pp while diplomacy communities nearly doubled (24.6\% $\rightarrow$ 42.5\%). Following Hanoi, humanitarian and economic communities expanded, suggesting agenda broadening rather than simple reversion. The predominance of neutral discourse in comments highlights that the observed shifts are not driven by wholesale opinion change, but by selective reorganization of how salient frames are connected within public discussion.

\textbf{Implications for Policy Communication: } While our findings are based on engaged online communities rather than representative samples, they suggest potential implications for diplomatic communication strategies: high-profile diplomatic initiatives may rapidly alter interpretive frames, even in adversarial contexts. Positive framing exhibits persistence following successful engagement (ratchet effect), and framing effects appear more pronounced than sentiment effects. Failed diplomacy does not fully undo earlier gains, indicating some resilience in discourse shifts. However, these patterns require validation across broader populations before informing policy directly.

\textbf{Limitations \& Future Work.}
Several limitations warrant consideration. First, our analysis is limited to English-language Reddit posts, which may not capture discourse dynamics on non-English platforms or in regions with minimal Reddit usage. Furthermore, Reddit users, who skew younger and more politically engaged, are not representative of the general public.
Second, while we employed a robust multi-control design, Russia's exclusion from the P1$\rightarrow$P2 framing analysis due to parallel trends violation ($p=0.01$) reduces our control group to two countries for the Singapore period. This limitation affects the robustness of our counterfactual estimates, as confounding control relies on having multiple comparison groups. 
Third, our aggregate analysis does not distinguish whether rhetorical shifts stem from individual belief changes or platform compositional churn (i.e., new users entering the conversation). Future work should employ user-level longitudinal models to disentangle these mechanisms.
Fourth, our 10-month post-Hanoi window captures \textit{short-term} asymmetric persistence; examining long-term durability against subsequent shocks (e.g., COVID-19) remains an open question.
Finally, our causal analysis focuses on original posts; while we include a supplementary comment-level extension, we do not model comment threads as independent causal units. Future research should extend this framework to comment threads to capture deeper deliberative dynamics.

\textbf{Ethical Considerations \& Broader Impact.} This study uses publicly available Reddit data accessed through the Arctic Shift API~\cite{arcticshift2024}. No personally identifiable information was collected, and all posts analyzed were publicly shared by users. While analyzing discourse dynamics supports transparent policymaking, we acknowledge potential risks: automated framing tools could be repurposed for computational propaganda. We advocate for their responsible use to detect rather than generate manipulative narratives.

%% file: sections/conclusion.tex
\section{Conclusion}
This study provides causal and structural evidence of how high-stakes diplomatic summits affect public discourse on social media. Using a Difference-in-Difference (DiD) design with LLM-based classification and expert-validated benchmarks, we find three key results. First, the 2018 Singapore Summit produced significant positive shifts in framing toward North Korea, with effect sizes substantially larger for framing than sentiment. Second, diplomatic engagement reorganized discourse networks, shifting both edge-level framing (threat $\rightarrow$ diplomacy) and community-level thematic structure. Third, the 2019 Hanoi Summit failure triggered only partial reversals; discourse did not return to pre-summit baselines, suggesting asymmetry in diplomatic effects. Importantly, these patterns propagated to highly visible audience responses. While comment-level emotional sentiment exhibited the same transience observed at the post level (fully reverting after Hanoi), the structural framing and network configuration of audience discourse maintained the asymmetric persistence observed in agenda-setting posts, confirming that top-down diplomatic signals can reshape bottom-up narrative structures even when emotional valence is fleeting.

%% file: sections/appendix.tex
\appendix
\section*{Appendix}
\setcounter{secnumdepth}{2}
\section{LLM Classification Prompt}
\label{app:llm_prompt}

The classification was performed using GPT-4o-mini with the following system and user prompts.

\subsection*{System Prompt}
You are a political science researcher analyzing media framing of international relations. Apply the Critical Classification Rules FIRST before classifying.

\subsection*{User Prompt}
You are an international relations researcher. Classify the following Reddit post into ONE of 5 framing categories.

\subsubsection*{Critical Classification Rules (Apply First!)}
These rules resolve ambiguities and edge cases.

\begin{enumerate}
    \item \textbf{No Action = NEUTRAL}: If the post is a question, hypothesis, speculation, or factual report without explicit government action, classify as NEUTRAL.
    \item \textbf{Verbal vs. Physical Actions}: If a state is \textbf{only verbally criticizing or warning} another state (not taking physical/military action), classify as DIPLOMACY, not THREAT.
    \item \textbf{Individual Harm = HUMANITARIAN}: If the harm is to \textbf{specific individuals} (protesters, defectors, refugees, civilians), classify as HUMANITARIAN, not THREAT.
    \item \textbf{Conflicting Frames = NEUTRAL}: When DIPLOMACY and THREAT (or other frames) are equally present and competing, classify as NEUTRAL.
    \item \textbf{Domestic Politics = NEUTRAL}: Commentary on domestic political issues, even if mentioning foreign countries, is NEUTRAL.
\end{enumerate}

\subsubsection*{Classification Criteria}

\paragraph{1. THREAT (Military Tension/Conflict)}
Physical military actions that increase conflict possibility.
\begin{itemize}
    \item \textbf{Include}: Military actions (missile launches, nuclear tests, military exercises, shows of force), arms buildup, military threats with NO dialogue possibility (ultimatums), cyberattacks on military/government infrastructure.
    \item \textbf{Exclude (classify as DIPLOMACY instead)}: Verbal warnings with possibility of dialogue remaining, one state verbally criticizing another's actions, requests to stop military activities.
\end{itemize}

\paragraph{2. DIPLOMACY (Diplomatic Interaction)}
Relationship adjustment through dialogue, negotiation, or verbal pressure.
\begin{itemize}
    \item \textbf{Include}: Summit meetings, diplomatic negotiations, bilateral/multilateral talks, agreements, attempts to improve/normalize relations, sanctions relief, verbal criticism/warnings between states, diplomatic pressure.
    \item \textbf{Note}: Even if a summit fails, focus on the summit itself $\rightarrow$ DIPLOMACY.
\end{itemize}

\paragraph{3. ECONOMIC (Economic Measures)}
Pressure or cooperation through economic means.
\begin{itemize}
    \item \textbf{Include}: Imposition/strengthening of economic sanctions, sanctions evasion activities, trade measures (tariffs), economic cooperation/investment.
    \item \textbf{Exclude}: Arms deals $\rightarrow$ THREAT, if main focus is diplomatic action $\rightarrow$ DIPLOMACY.
\end{itemize}

\paragraph{4. HUMANITARIAN (Humanitarian Issues)}
Human rights and individual/civilian harm.
\begin{itemize}
    \item \textbf{Include}: Human rights violations, refugee issues, humanitarian assistance/aid, war crimes, harm to individuals (protesters, defectors, refugees, civilians), cyberattacks targeting civilians.
\end{itemize}

\paragraph{5. NEUTRAL (Neutral Information)}
Cases not fitting specific frames.
\begin{itemize}
    \item \textbf{Include}: Simple factual reporting, analysis, domestic politics, questions/hypotheticals, description without explicit government action, when multiple frames are equally present.
\end{itemize}

\section{Parallel Trends Validation}
\label{app:parallel_trends}

Tables~\ref{tab:app_pt_sentiment} and~\ref{tab:app_pt_framing} present the full parallel trends test results for sentiment and framing outcomes, respectively. For sentiment, all three control groups satisfied parallel trends across both comparisons ($p > 0.05$). For framing, China and Iran satisfied parallel trends in all comparisons, while Russia exhibited a significant violation for P1$\rightarrow$P2 ($p = 0.01$), likely due to the Mueller investigation's confounding effect on threat-oriented discourse during this period. Consequently, Russia is excluded from framing analyses for the Singapore Summit effect.

\begin{table}[h]
\centering
\caption{Post Parallel Test Results (Sentiment)}
\small
\begin{tabular}{llcc}
\toprule
\textbf{Comparison} & \textbf{Control} & \textbf{P-value} & \textbf{Trends} \\
\midrule
P1$\rightarrow$P2 & China & 0.99 & Valid \\
P1$\rightarrow$P2 & Iran & 0.73 & Valid \\
P1$\rightarrow$P2 & Russia & 0.14 & Valid \\
\addlinespace
P2$\rightarrow$P3 & China & 0.71 & Valid \\
P2$\rightarrow$P3 & Iran & 0.64 & Valid \\
P2$\rightarrow$P3 & Russia & 0.81 & Valid \\
\bottomrule
\end{tabular}
\label{tab:app_pt_sentiment}
\end{table}

\begin{table}[h]
\centering
\caption{Post Parallel Test Results (Framing)}
\small
\begin{tabular}{llcc}
\toprule
\textbf{Comparison} & \textbf{Control} & \textbf{P-value} & \textbf{Trends} \\
\midrule
P1$\rightarrow$P2 & China & 0.06 & Valid \\
P1$\rightarrow$P2 & Iran & 0.69 & Valid \\
P1$\rightarrow$P2 & Russia & 0.01 & Invalid \\
\addlinespace
P2$\rightarrow$P3 & China & 0.40 & Valid \\
P2$\rightarrow$P3 & Iran & 0.76 & Valid \\
P2$\rightarrow$P3 & Russia & 0.30 & Valid \\
\bottomrule
\end{tabular}
\label{tab:app_pt_framing}
\end{table}

\section{Robustness Check: Alternative Framing Scale}
\label{app:robustness}

To verify that our main findings are not artifacts of the framing scale construction, we conducted a robustness check using an alternative specification that treats ECONOMIC framing as partially coercive (ECONOMIC = $-1$) rather than neutral (ECONOMIC = $0$).
This alternative reflects a view commonly adopted in international relations scholarship, where economic sanctions are conceptualized as coercive instruments situated between military threat and diplomatic engagement.

\begin{table}[h!]
\centering
\caption{DiD Estimates: Original vs. Alternative Framing Scale}
\resizebox{\linewidth}{!}{
\begin{tabular}{llcc}
\toprule
\textbf{Event} & \textbf{Control} & \textbf{Original} & \textbf{Alternative} \\
 & & (ECON=0) & (ECON=$-1$) \\
\midrule
Singapore & China & +0.34** & +0.37** \\
(P1$\rightarrow$P2) & Iran & +0.39 & +0.77** \\
\midrule
Hanoi & China & -0.34 & -0.23 \\
(P2$\rightarrow$P3) & Iran & -0.23 & -0.22 \\
 & Russia & -0.35 & -0.36 \\
\bottomrule
\multicolumn{4}{l}{\footnotesize * $p<0.05$, ** $p<0.01$, *** $p<0.001$} \\
\end{tabular}
}
\label{tab:app_robustness}
\end{table}

The alternative specification produces similar directional effects across all comparisons, with several estimates increasing in magnitude under the alternative scaling.
Notably, the Singapore/Iran estimate becomes larger and reaches statistical significance under the alternative specification (+0.39 $\rightarrow$ +0.77**), indicating that the baseline coding does not inflate diplomatic effects.
Taken together, these results suggest that our main specification (ECONOMIC = $0$) yields more conservative effect size estimates, while preserving both the direction and substantive interpretation of the findings.

\section{Audience Response Analysis (RQ3)}
\label{app:audience_response}

This appendix provides detailed statistics and results from the supplementary analysis of audience responses (comments), which serves to validate the propagation of framing effects beyond agenda-setting posts.

\subsection{Parallel Trends Verification}
To validate the DiD design for audience responses, we verified the parallel trends assumption for both sentiment and framing (Tables~\ref{tab:app_comment_pt_sentiment} and~\ref{tab:app_comment_pt_framing}). Unlike the post-level analysis where Russia violated parallel trends for framing, the audience response data satisfied the parallel trends assumption across \textit{all} control groups (Russia $p=0.615$), allowing us to include all three countries in the analysis.

\begin{table}[h]
\centering
\caption{Comment Parallel Test Results (Sentiment)}
\begin{tabular}{llcl}
\toprule
\textbf{Comparison} & \textbf{Control} & \textbf{P-value} & \textbf{Verdict} \\
\midrule
P1$\rightarrow$P2 & China & 0.51 & Valid \\
P1$\rightarrow$P2 & Iran & 0.20 & Valid \\
P1$\rightarrow$P2 & Russia & 0.61 & Valid \\
\addlinespace
P2$\rightarrow$P3 & China & 0.33 & Valid \\
P2$\rightarrow$P3 & Iran & 0.39 & Valid \\
P2$\rightarrow$P3 & Russia & 0.40 & Valid \\
\bottomrule
\end{tabular}
\label{tab:app_comment_pt_sentiment}
\end{table}

\begin{table}[h]
\centering
\caption{Comment Parallel Test Results (Framing)}
\begin{tabular}{llcl}
\toprule
\textbf{Comparison} & \textbf{Control} & \textbf{P-value} & \textbf{Verdict} \\
\midrule
P1$\rightarrow$P2 & China & 0.57 & Valid \\
P1$\rightarrow$P2 & Iran & 0.34 & Valid \\
P1$\rightarrow$P2 & Russia & 0.62 & Valid \\
\addlinespace
P2$\rightarrow$P3 & China & 0.32 & Valid \\
P2$\rightarrow$P3 & Iran & 0.92 & Valid \\
P2$\rightarrow$P3 & Russia & 0.50 & Valid \\
\bottomrule
\end{tabular}
\label{tab:app_comment_pt_framing}
\end{table}

\subsection{Structuring of Discourse: Communities and Edges}
\label{stru-dis}
Table~\ref{tab:app_comment_structure} details the distribution of framing within comment-derived knowledge graphs. The analysis reveals a persistent structural shift:
\begin{itemize}
    \item \textbf{Community Level}: DIPLOMACY-oriented communities doubled in P2 (28.6\%) and remained elevated in P3 (17.5\%). Notably, HUMANITARIAN communities surged in P3 (10.8\%), indicating a shift to human rights issues rather than returning to pure security threats.
    \item \textbf{Edge Level}: Threat-oriented relationships collapsed in P2 (-16.9 pp) and only partially recovered in P3, while DIPLOMACY edges remained elevated relative to pre-summit levels (P1: 15.9\% $\rightarrow$ P3: 19.7\%), confirming the structural persistence of engagement narratives.
\end{itemize}

\section{Cross-LLM Validation}
\label{app:cross_llm}

To ensure that our framing classifications are robust across different LLM families and model sizes, we validated our primary classifier (GPT-4o-mini) against multiple alternative models. GPT-4o-mini was selected as the primary classifier based on its strong agreement with human gold labels (85.0\% accuracy, Cohen's $\kappa$ = 0.79, $N=500$). We then tested cross-model consistency by comparing other LLMs against GPT-4o-mini predictions on a stratified sample of 500 posts.
\begin{table}[h!]
\centering
\caption{Cross-LLM Validation: Agreement with GPT-4o-mini ($N=500$)}
\resizebox{\linewidth}{!}{
\begin{tabular}{lcc}
\toprule
\textbf{Model} & \textbf{Agreement (\%)} & \textbf{Cohen's $\kappa$} \\
\midrule
GPT-4o & 85.7\% & 0.78 \\
Claude Sonnet 4.5 & 84.0\% & 0.76 \\
Claude Haiku 4.5 & 81.6\% & 0.73 \\
Llama 3.3 70B & 81.5\% & 0.71 \\
Llama 3.1 8B & 75.3\% & 0.62 \\
\bottomrule
\multicolumn{3}{l}{\footnotesize Primary: GPT-4o-mini (85.0\% accuracy vs. human labels, $\kappa$=0.79)} \\
\end{tabular}
}
\label{tab:cross_llm}
\end{table}

Table~\ref{tab:cross_llm} presents the agreement metrics. GPT-4o achieved the highest agreement at 85.7\% ($\kappa$ = 0.78), followed closely by Claude Sonnet 4.5 at 84.0\% ($\kappa$ = 0.76), Claude Haiku 4.5 at 81.6\% ($\kappa$ = 0.73), and Llama 3.3 70B at 81.5\% ($\kappa$ = 0.71). This high cross-family consistency, particularly between models from different developers and architectures (OpenAI, Anthropic, and Meta), provides strong evidence that our classification results reflect objective discourse patterns rather than model-specific artifacts.

\subsection{Qualitative Error Analysis}
\label{app:error_analysis}
Table~\ref{tab:app_error_examples} provides qualitative examples of classification disagreements between human annotators and the LLM. The analysis highlights two primary sources of error: (1) Over-application of the ``factual reporting = NEUTRAL'' heuristic to diplomatic events, and (2) Over-sensitivity of the ``individual harm = HUMANITARIAN'' rule to keywords like ``protesters'' or ``defectors'' even when the primary framing is not humanitarian.

\begin{table*}[h]
\centering
\small
\caption{Examples of Human-LLM Classification Disagreements}
\begin{tabularx}{\textwidth}{>{\raggedright\arraybackslash}p{4.8cm} c c >{\raggedright\arraybackslash}X}
\toprule
\textbf{Post Title} & \textbf{Human} & \textbf{LLM} & \textbf{LLM Reasoning (Summary)} \\
\midrule
\multicolumn{4}{l}{\textit{Type 1: Factual Reporting vs. Diplomatic Action}} \\
\addlinespace[0.3em]
Trump calls off planned Singapore summit & DIP & NEU & Classified as a ``factual report'' about an event cancellation without explicit government action. \\
\addlinespace[0.5em]
Kim Jong Un received an `excellent' letter from Trump & DIP & NEU & Viewed as a report about a letter rather than a significant diplomatic interaction. \\
\addlinespace[0.5em]
Tillerson in Beijing set to talk on North Korea & DIP & NEU & Interpreted as a report on travel plans rather than a substantive diplomatic engagement. \\
\addlinespace[0.5em]
North Korea cuts ICBMs out of military parade & DIP & NEU & Failed to recognize the parade modification as a deliberate diplomatic signal of de-escalation. \\
\addlinespace[0.5em]
South Korea's Leader Credits Trump for North Korea Talks & DIP & NEU & Characterized as a neutral report on a leader's statement rather than an act of diplomatic recognition. \\
\midrule
\multicolumn{4}{l}{\textit{Type 2: Keyword Sensitivity (Individual Harm)}} \\
\addlinespace[0.3em]
Anonymous Hacks China... Students Trapped & THR & HUM & Focus on ``Students trapped'' outweighed the military/cyber threat context. \\
\addlinespace[0.5em]
Russia delivered 12.1 metric tons aid to Syria & DIP & HUM & Interpreted ``aid delivery'' as a humanitarian act rather than a state-led diplomatic move. \\
\addlinespace[0.5em]
Facebook Blocks \$200K Donation To Iran Quake Victims Because Sanctions & NEU & HUM & Focus on ``Quake Victims'' overrode the primary context of corporate compliance with sanctions. \\
\addlinespace[0.5em]
North Korea defectors launch 'birthday' balloons across DMZ & NEU & HUM & The keyword ``defectors'' triggered a Humanitarian classification despite the political nature of the act. \\
\addlinespace[0.5em]
Otto Warmbier: Parents of man tortured in North Korea condemn Trump & NEU & HUM & The mention of ``tortured'' led to a Humanitarian label, missing the domestic political reaction frame. \\
\bottomrule
\end{tabularx}
\label{tab:app_error_examples}
\end{table*}

\section{Temporal Trend Visualizations}
\label{app:trend_figures}

\begin{figure*}[h!]
\centering
\includegraphics[width=0.8\textwidth]{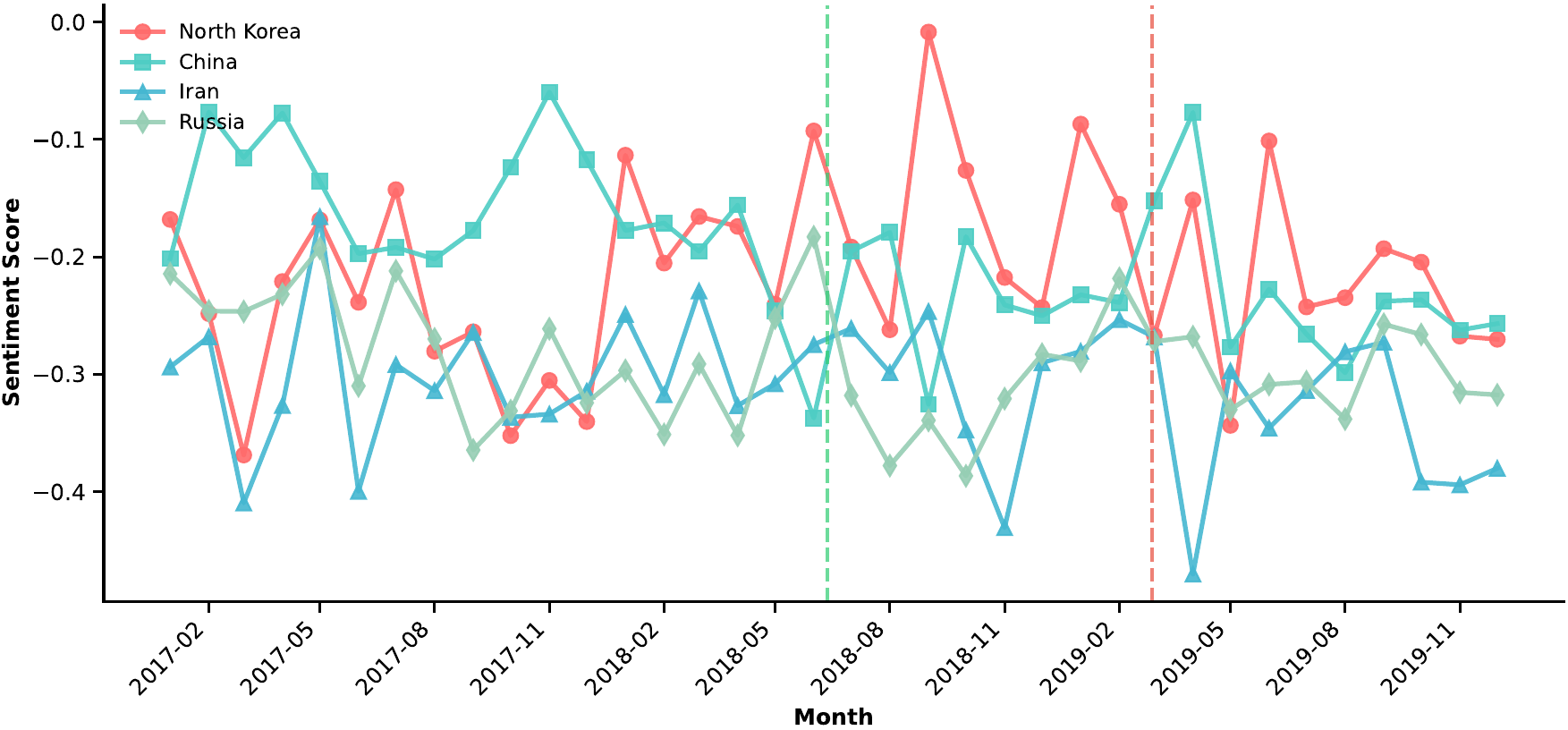}
\caption{Monthly Sentiment Score Trends: NK vs. Control Groups. Vertical dashed lines mark key diplomatic events: the green line indicates the Singapore Summit (June 2018), and the red line indicates the Hanoi Summit (February 2019).}
\label{fig:app_sentiment_trends}
\end{figure*}

\begin{figure*}[h!]
\centering
\includegraphics[width=0.8\textwidth]{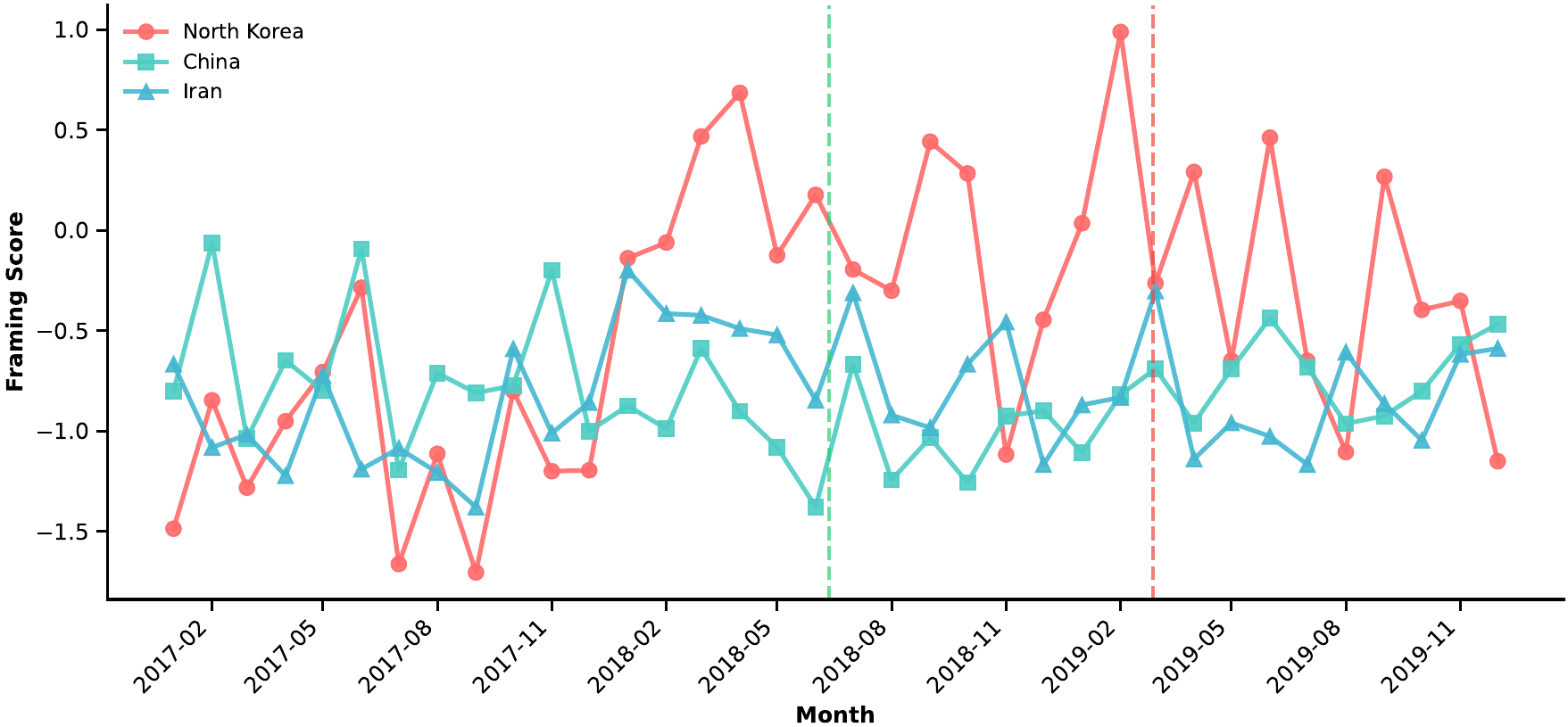}
\caption{Monthly Framing Score Trends: NK vs. Control Groups. Vertical dashed lines mark key diplomatic events: the green line indicates the Singapore Summit (June 2018), and the red line indicates the Hanoi Summit (February 2019).}
\label{fig:app_framing_trends}
\end{figure*}

Figures~\ref{fig:app_sentiment_trends} and~\ref{fig:app_framing_trends} present the monthly time-series trends of sentiment and framing scores across all countries in our study. These visualizations provide additional context for the Difference-in-Differences analyses presented in the main text, allowing visual inspection of pre-treatment parallel trends and post-treatment divergence.

